\documentclass[11pt]{amsproc}
 \usepackage{mathrsfs}

\usepackage{color}

\usepackage[shortlabels]{enumitem}

\def\XXint#1#2#3{{\setbox0=\hbox{$#1{#2#3}{\int}$ }
\vcenter{\hbox{$#2#3$ }}\kern-.6\wd0}}

\def \<{\langle}
\def \>{\rangle}

\usepackage{amsfonts}
\usepackage{amsmath,amsthm,amscd}
\usepackage{graphicx}
\usepackage{subfigure}
\usepackage[abbrev]{amsrefs}
\newtheorem{definition}{Definition}
\newtheorem{lemma}{Lemma}
\newtheorem{corollary}{Corollary}

\newcommand{\R}{\mathbb R}

\renewcommand{\phi}{\varphi}
\newcommand{\p}{\partial}
\newcommand{\w}{\omega}
\newcommand{\n}{\nabla}

 \def \lrc{\lrcorner}
 \def \V{{\mathcal V}}
 \def \kf{\mathfrak k}
 \def \V{{\mathcal V}}
 \def \G{{\mathcal G}}

\keywords{Yang-Mills, heat equation, weakly parabolic, gauge groups, Gaffney-Friedrichs inequality, Neumann domination.}

\title[The Yang-Mills heat equation]{The Yang-Mills heat equation on three-manifolds with boundary}

\author{Nelia Charalambous}
\address{Department of Mathematics and Statistics, University of Cyprus, Nicosia, 1678, Cyprus} \email[Nelia Charalambous]{nelia@ucy.ac.cy}

\dedicatory{This paper is dedicated to Leonard Gross}

\date{\today}

\begin{document}
\begin{abstract} In this short note we provide an expository account of the work of Leonard Gross and the author on the Yang-Mills heat equation over smooth three-manifolds with boundary.
\end{abstract}

\maketitle

\newtheorem{theorem}{Theorem}[section]
\newtheorem{proposition}{Proposition}[section]
\newtheorem{lem}{Lemma}[section]
\newtheorem{rem}{Remark}[section]

\def\query#1{\setlength\marginparwidth{55pt}%
\marginpar{\raggedright\fontsize{7.81}{9}\selectfont\itshape\hrule\smallskip
\color{red}#1\color{black}\par\smallskip\hrule}}
\def\removequeries{\def\query##1{}}

\section{Introduction}

A theory to explain why the nucleus of an atom does not fall apart, in spite of the powerful repelling
electric force between protons, was proposed by C.N.  Yang and R.L. Mills in 1954 \cite{YM54}. After quantization and systematic elimination of the divergences in the approximation schemes needed for the computation of experimental predictions, the theory is now widely regarded as
one of the most successful theories of fundamental physics. But the internal consistency of
the theory remains in question: It is not clear what the approximations are approximating.
After 70 years of intense efforts by many mathematicians and physicists to find the
mathematical structures  into which the computations fit, a solution does not yet seem to be in sight, in spite of the many
different approaches that have been explored.

The aim of our work is to understand the configuration space for a Yang-Mills field and,
by completion, to find an infinite dimensional support space for the presumed ground state measure,
which would help to give meaning to the approximations.  Informally, the configuration space
for a Yang-Mills field is a quotient space  $\mathcal{A}/\mathcal{G}$ where $\mathcal{A}$ is some (yet to be determined)
space of 1-forms on $\mathbb{R}^3$ with values in the Lie algebra of some compact
Lie group $K$. The Lie group $K$ is determined experimentally, and for example $K = SU(3)$ is now regarded as the correct group for strong interactions in the nucleus of a particle. $\mathcal{G}$ is some (yet to be determined) set of functions $g:\mathbb{R}^3 \to K$, which is a group under pointwise multiplication and which acts naturally on $\mathcal{A}$. The case $K = U(1)$ has provided further understanding of what the configuration space should be, and it is now clear that the very large completion needed to support the ground state measure must contain distributions over $\mathbb{R}^3$ of large negative Sobolev index.

Various classes of distributions over  $\mathbb{R}^3$  can be characterized by the behavior  of the solution to the standard heat equation on $\R^3$ whose initial value is a distribution. See for example \cite{Lio} and \cite{Tai1} for some  classical developments of this identification. For our purposes  the classical heat equation must be replaced by the Yang-Mills heat equation in order for  the gauge group to commute with the flow of the solution.
The Yang-Mills heat equation was first used  by D. Zwanziger \cite{Z} over $\mathbb{R}^4$
as a zeroth order approximation
 of stochastic quantization. Independently,  S.K. Donaldson \cite{Do1} introduced
the Yang-Mills heat equation over a complex surface (4 real dimensions) as a tool to study the existence of irreducible Yang-Mills connections on the projective plane. He demonstrated that a vector bundle on a complex surface is stable if and only if it admits a Hermitian Yang-Mills connection.

In our work we have been considering the Yang-Mills heat equation in  a product bundle over a compact 3-dimensional Riemannian manifold $M$ with possibly nonempty smooth boundary.  We consider a $K$-product-bundle over $M$, where $K$ is a compact, possibly nonabelian, connected Lie group with Lie algebra $\mathfrak{k}$. A connection over a product bundle allows us to take derivatives of its various differential structures, and in general there exist multiple options for a connection depending on the properties of the space that we would like to preserve. In the case of the tangent bundle for example, the  Levi-Civita  connection is the unique metric-preserving connection with zero torsion on tangent vectors. Over a $K$-product-bundle, we would also like to differentiate the structures that are generated by the Lie group and the Lie algebra, and differentiation `should' be invariant under a change of variables on the manifold, but also under the action of the Lie group on the bundle fibers. Such derivatives are given by the space of connections, which is an affine space over $\Lambda^1(\mathfrak{k})$, the set of 1-forms with values in the Lie algebra $\mathfrak{k}$.  The sections of the Lie group $K$ form the group of symmetries of the bundle, known as the gauge transformations.

The energy of a connection, also referred to as its magnetic energy in dimension 3, is given by the  Yang-Mills functional
\begin{equation}
\mathcal{YM}(A)= \int_M |B|^2\, dx
\end{equation}
where  $\;B:= dA  + A \wedge A \; $ is   the curvature of $A$ and $dx$ denotes the Riemannian volume element of the manifold. The variational equation for extrema of the Yang-Mills functional is the (degenerate elliptic) Yang-Mills equation $d^*_A B=0$, where  $d_{A}^*$ is the gauge covariant coderivative. On compact manifolds of dimension up to 4 existence and regularity of solutions to the Yang-Mills equation has been extensively studied by C. H. Taubes    and K. Uhlenbeck  whereas T. Isobe and A. Marini studied the case of   nonempty boundary  (see \cite{Ma7,Ma5,Ma3,Uh2,Uh1,Tau1,Tau2}). Taking the negative gradient flow corresponding to the Yang-Mills  functional gives rise to the (degenerate parabolic)   Yang-Mills heat equation.

The  Yang-Mills heat equation is given by
\begin{equation} \label{ymh}
 \p A(t)/ \p t = -d_{A(t)}^* B(t), \ \ \text{for} \ \  t >0\ \ \text{with}\ \    A(0) = A_0,
\end{equation}
for some adequate initial condition $A_0$, and where $B(t)$ denotes the curvature of $A(t)$. The major difficulty when studying its solutions is the fact that the second order operator on the right side is not fully elliptic. In fact, its solutions are invariant under the infinite group of gauge transformations and as a result the heat operator does not smooth out all initial data. This can be illustrated by the following simple example. Consider the connection  $A(t)= g^{-1} dg$ where $g$ is a time-independent gauge.  $A(t)$ has curvature $B(t)=0$, therefore $A(t)= g^{-1} dg$  is a solution to \eqref{ymh} with initial condition  $A(0)=  g^{-1} dg$. In other words, $A(t)$ can be as irregular as $A(0)$, even if the curvature $B(t)$ is smooth. In addition, \eqref{ymh} is nonlinear in the nonabelian case, as  a cubic term in $A$ appears in the right side. The above obstructions have limited results on the regularity and uniqueness of strong solutions to manifolds of dimension less than or equal to 3, due to the dimension restrictions of Sobolev embedding theorems.

Over compact manifolds with empty boundary in dimensions 2 and 3, J. Rade \cite{Rade} proved existence and uniqueness of strong solutions for initial data in $H_1$. Rade also demonstrated that the solution converges in  $H_1$ to a Yang-Mills connection as  $t\to \infty$, and that a Yang-Mills connection  for any energy $\lambda$ may be realized as a limit.  His method of proof constituted in adding a parabolic equation that the curvature $B(t)$ satisfies  to \eqref{ymh} and then solving the system. This technique is known as DeTurck's trick who first used it in the context of the parabolic Ricci flow \cite{DT}, and the method was also used by J. Ginibre and G. Velo in the context of the hyperbolic Yang-Mills equation \cite{GV1,GV2}.

On the other hand, it is well known that singularities may develop at finite time for dimensions 4 or higher, unless one assumes some strong symmetric properties for the solutions \cites{Do1,HT1,Stru2,SST}. Over Euclidean space $\mathbb{R}^n$ for $n\geq 5$ J. Grotowski showed that solutions can blow up at finite time even for smooth initial data \cite{Grot}. For the nature of the singularity formation see for example \cites{Ga,Na,We}. In the context of weak solutions M. Struwe, A.   Schlatter and A.  Tahvildar \cites{Stru,SST}  prove the existence of weak solutions (with finitely many isolated singularities) in dimension 4, whereas over compact 4-manifolds  A. Waldron  recently showed that smooth solutions can be extended for all time \cite{Wal}.

In the case where the underlying space is a noncompact Riemannian manifold, and for initial data in $H_1$, L. Sadun \cite{Sa} proved the existence of solutions in dimension 3, although he did not provide any uniqueness nor regularity results (see also \cite{Ha} for the Yang-Mills Higgs flow).  M.-C. Hong and G. Tian   proved the existence of a smooth solution $A(t)\in C(\,[\tau,T]; H_1)$, but did not study its uniqueness properties. They then used this solution to find certain symmetric solutions to the Yang-Mills equation over $\mathbb{R}^4$ and then construct non-self-dual Yang-Mills connections over   $S^4$  \cites{Ho,HT2,HT1}.

We began our study of the Yang-Mills heat equation motivated by its potential application to the regularization of quantized Yang-Mills fields. It appears that a gauge invariant regularization method for Wilson loop variables will be necessary for the construction of quantized
Yang-Mills fields, as the standard methods for regularizing a quantum field are inapplicable to gauge fields, even though they have been successful in studying scalar field theories. At the same time, lattice regularization of Wilson loop functions of the gauge fields has been the only useful gauge invariant regularization procedure so far, but has not produced a continuum limit.

In this context, the Yang-Mills heat equation offers a way to regularize a large class of irregular connection forms $A_0$, by providing a gauge invariant and essentially smooth connection $A(t)$ corresponding to $A_0$, along which the Wilson loop functions will be well defined. In our work with L. Gross, we were interested in the existence and uniqueness of strong solutions to \eqref{ymh} over manifolds with boundary. Our focus on manifolds with boundary was due to the local nature of the Wilson loop variables, and the fact that local quantum field theory requires the use of locally defined observables such as $A_0$ and functions of $A_0$.  The existence problem requires assuming boundary conditions for the solution and the initial condition. Our Neumann and Dirichlet boundary conditions were the natural ones from the analytical point of view, but for the intended applications in quantum field theory the gauge invariant Marini boundary conditions will most likely prove to be the important ones.

The presence of boundary did not allow for the use of DeTurck's trick as in \cite{Rade} due to complications in the boundary conditions for the curvature. Instead our method was to use a symmetry breaking technique on the right side of the equation, in order to make it parabolic (see \eqref{ymp}) and was similar to the technique employed in \cite{Do1,Sa,Z}.  Our long-time existence followed a more classical path, compared to the semiprobabilistic methods used by A. Pulemotov in \cite{Pu}. A key element in our approach was the fact that we avoided the use of negative Sobolev spaces, which do not behave well under the heat kernel when the manifold has boundary. At the same time, our strict adherence to gauge invariant estimates was responsible for much of the novelty in our approach.

The parabolic equation \eqref{ymp}  used in our approach was also recently used by S.-J.Oh and S.-J. Oh and D. Tataru on $\mathbb{R}^3$ ($\mathbb{R}^4$ resp.) together with the hyperbolic one to give a novel way to study the Cauchy problem for the hyperbolic equation in 3+1 (4+1 resp.) space-time \cites{Oh,OT}. They considered $H_1$ initial data and showed that over $\mathbb{R}^4$ the solution is either global, or it will blow up to a soliton.  Under an additional $L^3$ assumption for the curvature over time, they proved that the solution will converge to a zero-curvature connection. They also used solutions to \eqref{ymh} to study the hyperbolic equation in 4+1 space-time.

We studied the Yang-Mills heat equation in two main cases. In Section \ref{S2} we will provide a summary of the existence and uniqueness, as well as regularity results that the author obtained together with L. Gross for initial  data in $H_1$, referred to as the finite energy case.  In dimension 3 however, the critical Sobolev exponent in the sense of scaling for the Yang-Mills heat equation is one half. This is due to the fact that the Sobolev $H_a(\mathbb{R}^3)$  norm  of a 1-form is invariant under the scaling $\vec{x}\to c \vec{x}$ for $\vec{x}\in \mathbb{R}^3$ if and only if $a=1/2$. As a result, one anticipates that the most general case in which the Cauchy problem would be solvable is for initial data $A_0 \in H_{1/2}(\mathbb{R}^3)$. Gross studied this case in the recent article \cite{Gr2}. We will give an overview of the results of Gross and the author for initial in  data in $H_a$, with $1/2\leq a<1$, under the assumption of finite action, in Section \ref{FA}. We will also provide an account of some recent results of L. Gross which are aimed at producing a Hilbert space structure to an appropriate configuration space for the set of solutions to the Yang-Mills heat equation in Section \ref{S4}.

{\bf Acknowledgment} {The author would like to thank Leonard Gross for introducing her to the Yang-Mills heat equation, for his support and teachings throughout the years, and also for a detailed historic account of the problem.}

\section{Technical Description}

We let $M$ be a smooth Riemannian manifold of dimension 3 with possibly nonempty smooth boundary $\p M$.   We study the Yang-Mills heat equation on a product bundle $M\times \V \rightarrow M$, where $\V$ is a finite dimensional real (resp. complex) vector space with an inner product. $K$ will denote a compact, possibly nonabelian, connected and orthogonal (resp. unitary) group of the space $End\ \V$, of operators on $\V$ to $\V$.  We denote the  Lie algebra of $K$ by $\mathfrak{k}$, which may be identified with a real subspace of $End\ \V$. We assume that we have an $Ad\ K$ invariant inner product $\langle \cdot, \cdot \rangle$ on $\mathfrak{k}$, with norm denoted by $|\cdot|_{\frak k}$.  We will not distinguish between  $|\xi|_{\frak k}$ and $|\xi|_{End \V}$, since they are equivalent norms and will usually denote this norm $|\cdot|$ for simplicity.  For $\mathfrak{k}$-valued $p$-forms $\omega, \phi$ we define $(\omega,\phi)=\int_M \langle \omega, \phi \rangle \, dx$ and the $L^2$ norm of a form by $\|\omega\|_2= (\omega,\omega)^{1/2}$. The $L^p$ norm is defined similarly for $1\leq p\leq \infty$ and denoted as $\|\omega\|_p$.

Over a product bundle, a connection can be identified with a  $\mathfrak{k}$-valued 1-form, and in local coordinates a connection may be written as $A=\sum_i A_i \, dx^i\ $ with $\ A_i \in \mathfrak{k}$.  The  curvature of $A$ is defined as
\[
B:= dA + (1/2)[A\wedge A]
\]
where $[A\wedge A]=\sum_{i,j}[A_i , A_j] \, dx^i\wedge dx^j$ and $[A_i , A_j]$ is the commutator in the Lie algebra. Each connection induces an exterior derivative $d_A: \Lambda^p(\mathfrak{k})\to \Lambda^{p+1}(\mathfrak{k})$,
such that for any $p$-form $\omega$
\[
d_A \omega = d\omega + [A \wedge \omega],
\]
and its adjoint $d_A^*: \Lambda^{p+1}(\mathfrak{k})\to \Lambda^{p}(\mathfrak{k})$
\[
d_A^* \omega = d^*\omega + [A \lrc \omega].
\]

A gauge transformation $g$ acts on a connection $A$ via
\[
A^{g} = g^{-1} dg + g^{-1}A\,g.
\]
$A, \;A^{g}$ are distinct matrix connections representing the same connection, and $g$ corresponds to a change in trivialization. Two connections are called  gauge equivalent, if they lie in the same orbit of this action. In this context, both the space of connections and the group of  gauge transformations are infinite dimensional sets.

We denote the gauge-covariant $W_1$ norm by
\[
\|\omega\|_{W_1}= \|\nabla \omega\|_2 + \|\omega\|_2
\]
which is defined independently of boundary conditions, and where $\nabla$ is the Riemannian covariant derivative on forms. Since our work is carried out over manifolds with boundary, we will distinguish the Sobolev space $W_1$, which does not assume any boundary conditions on the form $\omega$, from the Sobolev space $H_1$, and in general $H_a$, which does include boundary conditions for the form.





In our work we have been primarily interested in the existence and uniqueness of strong solutions to the Yang-Mills heat equation, which we define below.
\begin{definition} \label{SS1} Let $0 < T \leq \infty$. A {\bf \it  strong solution} to the Yang-Mills heat equation over $[0, T)$
is a continuous function
\begin{equation*}
A(\cdot): [0,T) \rightarrow  W_1(M; \Lambda^1 \otimes   \frak k )
\end{equation*}
such that
\begin{align}
a)& \ B(t):= dA(t)+ (1/2) [A(t)\wedge A(t)]  \in W_1 \  \text{for each}\ \  t\in (0,T),\          \notag     \\
b)& \  \text{the strong $L^2(M)$ derivative $A'(t) \equiv dA(t)/dt $}\
                                      \text{exists on}\ (0,T),   \text{and}       \notag \\
 &\ \ \ \ \ \  A'(\cdot): (0, T) \rightarrow L^2(M) \ \text{is continuous},    \notag  \\
c)& \  A'(t) = - d_{A(t)}^* B(t)\ \ \text{for each}\ t \in(0, T).        \notag
\end{align}
A solution $A(\cdot)$ that satisfies all of the above conditions
 except for $a)$ will be called an {\it almost strong solution}. In this
 case the spatial exterior derivative $dA(t)$, which appears in the
 definition of the curvature,  must  be interpreted in the weak sense.

A strong solution will be called   locally bounded if
\begin{align*}
&d) \  \| B(t)\|_\infty\ \text{is bounded on each
                             bounded interval $ [a,b) \subset (0, T)$ and} \\
&e) \  t^{3/4} \| B(t)\|_\infty\
           \text{is bounded on some interval $(0, b)$ with $0<b <T$.}
\end{align*}
\end{definition}

The above definition of a strong solution was the one used for problems where the initial condition has finite energy, in other words it belongs to $W_1$, with boundary conditions which we will discuss below. As will see in Section \ref{FA}, Gross was able to generalize the existence and uniqueness properties for less regular initial conditions, namely for $A_0$ in a Sobolev space $H_a(M)$ for $1/2\leq a < 1$. In this latter case since the initial condition does not belong to $W_1$, the solution itself cannot be continuous in $W_1$ at $t=0$, although the flow regularizes the initial condition for $t>0$. To clarify this distinction we will call these strong solutions of the second type, even though they are referred to as simply strong solutions in all of the literature.
\begin{definition} \label{SS2} Let $0 < T \leq \infty$. A {\bf \it  strong solution of the second type} to the Yang-Mills heat equation over $[0, T)$
is a continuous function
\begin{equation*}
A(\cdot): [0,T) \rightarrow  L^2(M; \Lambda^1 \otimes   \frak k )
\end{equation*}
such that
\begin{align}
a)& \ A(t) \in W_1 \ \text{for all} \ t \in (0,T) \ \text{and} \  A(\cdot): (0,T) \rightarrow W_1 \ \text{is continuous},  \notag \\
b)& \ B(t)  \in W_1 \  \text{for each}\ \  t\in (0,T),\          \notag     \\
c)& \  \text{the strong $L^2(M)$ derivative $A'(t) \equiv dA(t)/dt $}\
                                      \text{exists on}\ (0,T),   \text{and}       \notag \\
 &\ \ \ \ \ \  A'(\cdot): (0, T) \rightarrow L^2(M) \ \text{is continuous},    \notag  \\
d)& \  A'(t) = - d_{A(t)}^* B(t)\ \ \text{for each}\ t \in(0, T).        \notag
\end{align}
A solution $A(\cdot)$ that satisfies all of the above conditions
 except for $a)$ will be called an {\it almost strong solution of the second type}. In this
 case the spatial exterior derivative $dA(t)$, which appears in the
 definition of the curvature,  must again  be interpreted in the weak sense.
\end{definition}

If the boundary of the manifold is nonempty, then we must impose boundary conditions on the solutions.
\begin{definition}\label{defbdyconds}
For a strong solution to the Yang-Mills heat equation we will consider three types of boundary conditions:

\noindent
{\it Neumann boundary conditions:}
\begin{align}
&i)\ \ \  A(t)_{norm} =0\ \  \text{and}    \label{N1}\\
&ii)\ \  B(t)_{norm}=0\ \     \label{N2}
\end{align}
\noindent
{\it Dirichlet boundary conditions:}
\begin{align}
 &i)\ \ \  A(t)_{tan} =0\ \ \   \text{and}    \label{D1}\\
  &ii)\ \   B(t)_{tan} =0. \ \ \               \label{D2}
 \end{align}
 {\it Marini boundary conditions:}
\begin{align}
  &i)\ \ B(t)_{norm} =0. \ \ \             \label{M}
 \end{align}
 \end{definition}
The above Neumann and Dirichlet boundary conditions are related to the respective conditions for the Hodge Laplacian on differential forms but they are weaker, reflecting the weak parabolicity of the problem. The definition of tangential and normal components of a form on the boundary is the one generalized from the classical case of forms on manifolds with smooth boundary. For their precise definition as well as a further discussion of the Marini condition and our weaker boundary conditions see Section 2 in \cite{ChG1}. Here we would  simply like to remark that the Marini boundary condition is a nonlinear condition which is  gauge invariant.

As we have mentioned, the method of proof of our main results includes a symmetry breaking technique that replaces the original equation with a parabolic one where the Hodge Laplacian on 1-forms appears. We recall that
\begin{equation}
-\Delta = d^* d + d d^*,                                 \label{Lap1}
\end{equation}
where $d$ denotes the closed version of the exterior derivative operator
with $C_c^\infty(\R^3, \Lambda^1\otimes \kf)$ as a core. When the boundary of the manifold is nonempty there are many ways to define an adequate Sobolev space for the domain of this operator. The Sobolev spaces for  $\kf$ valued 1-forms that are associated to the boundary conditions that we considered can be obtained from the corresponding Laplacian. The classical Neumann and Dirichlet boundary conditions  are given by
\begin{align*}
&\w_{norm} =0  \ \ \text{and} \ \ (d\w)_{norm}=0, \ \  \ \text{\it Neumann conditions}    \\
&\w_{tan} =0\ \ \ \ \text{and}  \ \   (d^*\w)_{\p M} =0,  \  \ \  \text{\it Dirichlet conditions}.
\end{align*}
Alternatively, the Neumann  (resp. Dirichlet) Laplacian can be defined by \eqref{Lap1},
wherein $d$ is taken to be the maximal (resp. minimal) exterior derivative operator over $M$.
See \cite{ChG1} for further discussion of these domains. In both cases the Laplacian is
 a nonnegative, self-adjoint operator on the appropriate domain.

For $0 \le a \le 1$ we define the Sobolev spaces
\begin{equation}
H_a = \text{Domain of} \ (-\Delta)^{a/2} \ \text{on} \ L^2(M; \Lambda^1\otimes \mathfrak{k}) \notag
\end{equation}
with norm
\begin{equation}
\|\w\|_{H_a} = \|(1-\Delta)^{a/2} \w\|_{ L^2(M; \Lambda^1\otimes\mathfrak{k})}.  \label{Lap5}
\end{equation}
The following embedding property holds
\[
\|\w\|_{H_a} \leq c_{a,b} \|\w\|_{H_b} \ \text{whenever} \ 0\leq a \leq b,
\]
for some constant $c_{a,b}$ independent of $M$.

For solutions corresponding to $A_0\in H_a(M)$, one must define an appropriate group of gauge transformations that would work well when applying the symmetry breaking method for existence of solutions. For $a\in (1/2, 1]$ Gross defines the gauge group $\mathcal{G}_{1+a}$ which is in fact a Hilbert manifold \cite{Gr2}. For the case $a=1/2$ however, the corresponding group does not have a tangent space at the identity. This makes the analysis of the critical case $A_0\in H_{1/2}(M)$ all the more interesting. Below we give the full definition of $\mathcal{G}_{1+a}$ following \cite{Gr2}. In this case the underlying manifold $M$ is either $\mathbb{R}^3$ or the closure of a bounded open set in  $\mathbb{R}^3$ with smooth boundary.
\begin{definition}[The gauge group $\mathcal{G}_{1+a}$.] \label{defgg}   A measurable function
$g:M\rightarrow K \subset End\ \V$ is a bounded function into the linear space
$End\ \V$, therefore its weak derivatives are well defined. Following \cite{Gr2} we will write $g \in W_1(M;K)$
if $\|g - I_\V\|_2 <\infty$ and the derivatives $\p_j g \in L^2(M; End\ \V)$.
The 1-form $g^{-1} dg := \sum_{j=1}^3 g^{-1}(\p_j g)dx^j$ is then an a.e. defined  $\kf$ valued 1-form.
The Sobolev norm $\|g^{-1}dg \|_{H_a}$ is defined as in  \eqref{Lap5}.
For an element $g \in W_1(M;K)$
 the restriction $g|_{\p M}$ is well defined almost everywhere on $\p M$ by a Sobolev trace theorem.
The three versions of $\G_{1+a}$ that we will need are given in the following definitions.
\begin{align*}
\G_{1+a}(\R^3)
   = \Big\{g \in W_1(\R^3; K): g^{-1}dg \in H_{a}(\R^3;\Lambda^1\otimes \kf) \Big\}, \qquad \qquad \ \ \
\end{align*}
If $M \ne \R^3$ define
\begin{align*}
\G_{1+a}^N(M) &= \Big\{g \in W_1(M; K): g^{-1}dg \in H_{a}(M;\Lambda^1\otimes \kf) \Big\}, \\
\G_{1+a}^D(M) & = \Big\{g \in W_1(M; K): g^{-1}dg \in H_{a}(M;\Lambda^1\otimes \kf),\  g = I_\V\ \text{on}\ \p M \Big\}.
\end{align*}
It should be understood that the two spaces denoted  $H_{a}(M;\Lambda^1\otimes \kf)$ are those
determined by Neumann, respectively Dirichlet,  boundary conditions. It was proved in \cite[Theorem 5.3]{Gr2}
that all three versions of $\G_{1+a}$ are complete topological groups in the metric
$\rho_a(g,h) = \| g^{-1} dg - h^{-1} dh\|_{H_{a}} +\| g-h\|_{L^2(M; End\, \V)}$.
\end{definition}

The apriori energy estimates needed in our proofs must be in terms of gauge covariant derivatives, which reflect the many symmetries of solutions to the Yang-Mills heat equation, because neither the connection form nor its curvature is smoothed by the flow. As a result, it was necessary to express Sobolev inequalities in terms of the gauge covariant exterior derivative $d$ and its adjoint $d^*$. These will be elaborated on in Subsection \ref{Apr}.

\section{The Yang-Mills heat equation under finite energy} \label{S2}

In \cite{ChG1} we considered the Yang-Mills heat equation on 3-manifolds with smooth boundary when the initial condition $A_0$ is a connection lying in the first order Sobolev space $W_1(M)$, with an appropriate `half' boundary condition. We showed that there exists a unique  solution  to  \eqref{ymh} satisfying Dirichlet  or Neumann type boundary conditions. We also considered Marini boundary conditions where we proved existence and uniqueness for the flow whenever the initial data $A_0$  is $C^2$. Our main existence and uniqueness results in \cite{ChG1} are summarized below.
\begin{theorem} \label{thm1}
Suppose that $A_0 \in W_1$ and $(A_0)_{norm} =0.$ Then there exists a locally bounded strong solution $A(\cdot)$ over $[0, \infty)$ to \eqref{ymh}  such that $A(0) = A_0$,  which satisfies
the Neumann boundary conditions \eqref{N1}, \eqref{N2} as follows
\begin{equation} \label{N4}
A(t)_{norm} =0\ \ \text{for all}\ t \ge 0 \ \ \text{and}  \ \    B(t)_{norm}=0\ \   \text{for all}\  t >0.
\end{equation}
Moreover, if $A_1$ and $A_2$ are two locally bounded strong solutions which agree at $t=0$ and satisfy \eqref{N2} for $t>0$, then $A_1=A_2$ for all $t\in [0,\infty)$.

In the Dirichlet case, whenever $A_0 \in W_1$ and $(A_0)_{tan} =0,$ then there exists a locally bounded strong solution $A(\cdot)$ over $[0, \infty)$ to \eqref{ymh},  such that $A(0) = A_0$  which satisfies
the Dirichlet boundary conditions \eqref{D1}, \eqref{D2}
\begin{equation} \label{D4}
 A(t)_{tan} =0\ \ \ \text{for all}\ t \ge 0 \ \ \text{and}   \ \ B(t)_{tan} =0 \ \ \ \text{for all}\ t >0.
\end{equation}
If $A_1$ and $A_2$ are two locally bounded strong solutions which agree at $t=0$ and satisfy \eqref{D1} for all $t\geq0$, then $A_1=A_2$ for all $t\in [0,\infty)$.

For the case of Marini boundary conditions, whenever $A_0 \in C^2$ then there exists a unique locally bounded strong solution $A(\cdot)$ over $[0, \infty)$ to \eqref{ymh},  such that $A(0) = A_0$  which satisfies the Marini boundary condition
\begin{equation} \label{M2}
B(t)_{norm} =0 \ \ \ \text{for all}\ t > 0.
\end{equation}
\end{theorem}
Observe that the boundary condition for the uniqueness property is not symmetric; the Neumann problem only requires the boundary condition for the curvature, whereas the Dirichlet problem requires both, since $ A(t)_{tan} =0$ implies  $B(t)_{tan} =0$ when $B(t) \in W_1$. In the case of Marini boundary condition, the boundary condition itself suffices for the uniqueness result.

As Gross later observed in \cite{Gr2}*{Theorem 2.24} the above theorem will also hold in the case $M=\mathbb{R}^3$, as all the steps in our  proof will go through without any significant modifications since we never use that the volume of $M$ is finite.


Many of the  apriori  energy estimates that we required for the proof of the above theorems are usually formulated in terms of gauge covariant derivatives. In our case however, neither the connection form nor its curvature are smoothed by the flow. It was therefore necessary to express Sobolev inequalities in terms of the gauge covariant exterior derivative $d_A$ and its adjoint. We achieved this by proving a gauge invariant version of the Gaffney-Friedrichs inequality. The curvature of the connection form $A$ that appeared in these inequalities  contributed to some of the technical difficulties that needed to be resolved. However, our necessity to adhere to gauge invariant estimates was one of the innovative elements of our approach. It is also noteworthy that our method allowed us to obtain estimates for $\|A(t)\|_{W_1}$ although it is not a gauge invariant quantity.

\subsection{Existence by symmetry breaking}

The proof of the existence of solutions to the Yang-Mills heat equation relied on a symmetry breaking technique which consisted in adding a Zwanziger gauge fixing term $-d_A d^*A$  to the right side of \eqref{ymh}. To distinguish the solution to \eqref{ymh} from the solution to  the modified equation we will denote the latter by $C(t)$. The modified equation then becomes
\begin{equation} \label{ymp}
\p C(t)/ \p t = -d_{C(t)}^* B_C(t) - d_{C(t)}\, d^{*}C(t)
\end{equation}
where $B_C$ is the curvature of $C$. The Zwanziger term turns the second order operator on the right side into an elliptic one equal to $\Delta C + V(C)$  where $V$ is a nonlinear term of the type $V(C)= C^3 + C \cdot \partial C$. Although the solution to this modified  parabolic equation is no longer gauge invariant, it  be   transformed to a solution of the original equation using a time-dependent gauge transformation. This method was first proposed by D. Zwanziger \cite{Z} in the context of stochastic quantum field theory,  and a similar approach was used by S. K. Donaldson and separately L. Sadun \cites{Do1,Sa} in the context of the classical Yang-Mils heat equation. We refer to this method as the Zwanziger-Donaldson-Sadun (ZDS) procedure. In our work we had to be slightly more careful with our gauge fixing term due to the boundary conditions.

The parabolic equation does have a smooth unique solution for initial conditions in $W_1$, with boundary conditions given by
\begin{align}
&(N) \ \ C(t)_{norm}=0 \ \ \text{for} \ t\ge 0, \ \ \left(B_C(t)\right)_{norm}=0  \ \ \text{for} \ t>0 \label{N3}\\
&(D) \ \  C(t)_{tan}=0   \ \ \text{for} \ t\ge 0, \ \ d^*C (t)_{tan}= d^*C (t)\big|_{\p M}=0 \ \ \text{for} \ t>0. \label{D3}
\end{align}
These boundary conditions correspond the classical absolute (Neumann) and relative (Dirichlet) boundary conditions respectively, for real valued forms. We proved the following existence and uniqueness theorem for the parabolic equation \cite{ChG1}.
\begin{theorem}
Let $A_0 \in W_1$ satisfying either  $(A_0)_{norm} =0$ or respectively $ \,(A_0)_{tan} =0$. Then, there exists  $T>0$  and a continuous function $C:[0,\infty) \to W_1$  such that $C(0)=A_0$ and
\begin{align}
1)& \ B_C(t) \in W_1 \  \text{and} \ d^*C(t) \in W_1 \ \text{for each} \ t \in (0,T),  \notag \\
2)& \  \text{the strong $L^2(M)$ derivative $dC(t)/dt $}\
                                      \text{exists on for each}\ t\in (0,T),      \notag \\
3) &\  C(t) \ \text{satisfies} \  \eqref{ymp}   \ \text{together with the boundary conditions} \ \eqref{N3},  \ \text{respectively}\ \eqref{D3}, \notag \\
 &\  \ \ \text{for each} \ t \in (0,T),    \notag  \\
4)& \ t^{3/4}\|B_C(t)\|_{\infty} \  \text{is bounded on}\  \in(0, T).        \notag
\end{align}
The solution is unique under the above conditions. Moreover, $C(\cdot)$ belongs to $C^{\infty}(\,(0,\infty)\times M;\Lambda^1 \otimes   \frak k)$.
\end{theorem}

The proof of the above theorem was based on a classical conversion of the differential equation into an integral one, together with a contruction mapping argument into an appropriate Banach space that involved the $W_1$ norm of $C$ and the $L^\infty$ norms of $C, \; dC$ and $d^*C$. A regularity argument then allowed us to prove the higher order estimates. Uniqueness and the full boundary conditions for $C(t)$ follow in a similar way as in \cite{Tay3}, and the specific boundary conditions for the Yang-Mills problem follow from the symmetry properties of the operators involved.

For the smooth case, one can obtain the solution to the Yang-Mills heat equation from the parabolic equation \eqref{ymp} using the following gauging procedure.  For a smooth solution $C(t)$ to \eqref{ymp} with $C(0)=A_0$ that satisfies the boundary conditions \eqref{N3} (resp. \eqref{D3}), we define  the flow of gauge transformations $g(t):[0,\infty)\to C^{\infty}(M;K)$ as the solution to the initial value problem
\[
(\p g(t)/ \p t)  \;  g(t)^{-1}= d^{*} C(t), \ \ \  g(0) = I_K
\]
where $I_K$ is the identity element of the gauge group.  Then,
\[
{A}(t)= C(t)^{g(t)}=g(t)^{-1} dg(t) + g(t)^{-1} C(t) g(t)
\]
is a solution to \eqref{ymh} with $A(0)=A_0$ that satisfies the Neumann boundary conditions \eqref{N1} and \eqref{N2} (resp.  \eqref{D1} and \eqref{D2}) as in Theorem \ref{thm1}.

However, for initial data in $W_1$ singularity issues arise for $d^*C$ as $t\downarrow 0$ which make it difficult to obtain sufficient regularity for $g(t)$ so that $A(t) \in W_1(M)$ for $t\geq 0$. As we have mentioned, this is due to the fact that  $\|A(t)\|_{W_1}$  is not a gauge invariant quantity. In \cite{ChG1} we addressed this difficulty  by trying to avoid the singular point at $t=0$ for the gauge flow. In particular, we considered a solution $g_\epsilon (t)$ to the same equation as above, but only for $t\geq \epsilon$ and with initial condition $g_\epsilon (\epsilon)=I_K$.  Then for $t\geq \epsilon,$ $\,A_\epsilon(t) =C(t)^{g_\epsilon (t)}\,$  is a sequence of smooth solutions which strongly converges to a $W_1$ solution to the Yang-Mills heat equation solution as $\epsilon \downarrow 0$. The proof of this convergence was the most novel part of our work, and it relied on the gauge invariant Gaffney-Friedrichs inequality described in the following subsection.

\subsection{A priori estimates and a new  Gaffney-Friedrichs inequality} \label{Apr}

For any $\frak k$ valued $p$-form $\w$ on $M$ the covariant $W_1$ norm with respect to (a sufficiently smooth) connection $A$ is defined as
\[
\|\w\|_{W_1^A(M)}^2 = \|\n^A \w \|_2^2 + \| \w \|_2^2.
\]
where $\n^A$ is the covariant derivative induced by $A$. This gauge covariant norm is the one that can be used to control $L^p$ norms via Sobolev inequalities. For example, using the Sobolev and Kato inequalities one can show that
\begin{equation} \label{eqSob}
\|\w \|_6^2 \le  C(M) \|\w\|_{W_1^A(M)}^2 \
                          \   \text{for any}\  \w\  \text{and}\ A \in W_1(M)
                          \end{equation}
where $C(M)$ is a constant that depends only on the  geometry of $M$, but not on $A$. However, it is the Hodge version of the energy that relates well with the Yang-Mills equation
\[
\|d_A \w \|_2^2  + \|d_A^* \w \|_2^2 + \lambda \| \w\|_2^2
\]
due to the various symmetries that the solution and its curvature exhibit.

The  Gaffney-Friedrichs inequality that we proved in \cite{ChG1} is an important tool that allows us to relate the two and thus prove critical apriori estimates for solutions.

\begin{theorem}[Gaffney-Friedrichs inequality] \label{thmGF}
Let $M$ be a compact smooth 3-manifold with smooth boundary.  Suppose that $A\in W_1(M)$ and its curvature $B$ satisfies $\|B\|_2<\infty$. Then for any  $p$-form  $\w$   in $W_1(M)$ which satisfies either
\[
\w_{norm} =0 \ \ \ \text{or}\ \ \ \w_{tan} =0
\]
the following inequality holds
\[
(1/2) \|\w\|_{W_1^A(M)}^2\le  \|d_A \w \|_2^2  + \|d_A^* \w \|_2^2 + \lambda(B) \| \w\|_2^2
\]
where
\[
\lambda(B) :=  \lambda_M + \gamma_2 \|B\|_2^{4},
\]
and $\lambda_M$, $\gamma_2 $ depend  only on the geometry of $M$ and its boundary, but not on the size of $M$, neither on $A$.
\end{theorem}
Note that in \cite{ChG1} we also have a version of Theorem \ref{thmGF} for $B \in L^p$, for any  $p \in [2,\infty]$. Also, the constant $\lambda_M$ is zero in case the boundary of $M$ is convex.

Smooth solutions to \eqref{ymh} have a lot of symmetries. For example since $B$ satisfies the Bianchi identity $d_A B=0$, it follows that $B'= d_A A'$ and as a result the following differential inequality holds,
\begin{equation} \label{eqFE}
d/dt(\|B\|_2^2) = 2(B', B) = 2(d_A A', B)= -2 \|A'\|_2^2 \leq 0
\end{equation}
which in turn implies that the energy of a solution, $\|B(t)\|_2^2$, is nonincreasing with respect to $t$, and  therefore uniformly bounded by $\|B_0\|_2^2$.  At the same time, by combining equation \eqref{eqSob} and Theorem \ref{thmGF} (and under the appropriate boundary conditions) we can obtain an upper bound for the $L^6$, as well as the $W_1$ norm of the gauge invariant quantities $B$ and $A'$ since
\begin{equation*}
\begin{split}
\|B\|_6^2 &\leq c (\|d_A B\|_2^2 + \|d_A^* B\|_2^2 + \|B\|_2^2) \\
& =  c (\|A'\|_2^2 + \|B\|_2^2)
\end{split}
\end{equation*}  and
\begin{equation*} \begin{split}
\|A'\|_6^2 &\leq c (\|d_A A'\|_2^2 + \|d_A^* A'\|_2^2 + \|A'\|_2^2) \\
& =  c (\|B'\|_2^2 + \|A'\|_2^2)
\end{split}\end{equation*}
with respect to the $L^2$ norms $A, A'$ and $B$. Moreover we have the differential inequality
\begin{equation*}
d/dt(\|A'\|_2^2) \leq - \|B'\|_2^2 + c (\|B_0\|_2) \left[ \|A'\|_2^2  + \|A'\|_6^2 \right]
\end{equation*}
The above pointwise and integral identities can be used in combination with the Gaffney Friedrichs inequality to obtain  $L^6$ and $L^2$ bounds for $A'$ and $B$  with respect to the energy of the initial condition, $\|B_0\|_2$.  After some careful work and via interpolation arguments and the use of H\"older's inequality we can also use such estimates to prove that a smooth solution with $A_0\in W_1$ will remain in $W_1$ for all $t>0$ \cite{ChG1}*{Sections 5, 6}.

\subsection{Existence and Uniqueness}  The various apriori estimates obtained in the process outlined above, can be used to prove integrability estimates for the sequence of gauge transformations $g_\epsilon$. Ultimately they allowed us show that $A_\epsilon$ and $B_\epsilon$ are uniformly Cauchy in the $H_1$ norm as $\epsilon \downarrow 0$, and  also to prove all the regularity properties of a strong solution. Uniqueness over $[0,T)$ is a consequence of Gronwall's Lemma, because given our boundary conditions we can prove the inequality
\[
d/dt\|A_1(t) -A_2(t)\|_2^2\leq c(\|B_1(t)\|_{\infty}+\|B_2(t)\|_{\infty}) \, \|A_1(t) -A_2(t)\|_2^2
\]
where $B_1$ and  $B_2$ are the curvatures of $A_1$ and $A_2$ respectively.

For long time existence, the following regularization result was necessary.
\begin{lemma}   \label{lemReg}  Suppose that $A$ is a locally bounded strong solution
 over $[0, T)$ for some $T \leq \infty$. Let $ 0 < t <T$ and define
 $\beta = \sup_{0\le s \le t} \|A(s)\|_{W_1}$. Then there exists $\tau(\beta) >0$, such that, for any $[a, b] \subset (0, t]$
with $b-a <\tau$,  there exists a sequence $A_n$ of smooth solutions over  $[a,b]$ such that
 \begin{align*}
 \sup_{ a \le s \le b} \Big\{ \|A_n(s)& - A(s)\|_{W_1}
                                         + \|A_n'(s) - A'(s)\|_{L^2}                   \notag \\
 &+ \|B_n(s) - B(s)\|_{W_1}
    +\| B_n(s) - B(s)\|_\infty \Big\} \rightarrow 0
 \end{align*}
 as $n\rightarrow \infty$.
 \end{lemma}

Apart from long-time existence, this Lemma also allowed us  to prove further regularity properties for solutions. For example, given that the norms of $B$ and $A'$ are gauge invariant, we can prove the same $L^\infty$ bounds for the rough solution from the nearby smooth solutions, and therefore show that our strong solution to \eqref{ymh} is a locally bounded one.

\subsection{Neumann heat kernel domination} In \cite{ChG2} we continued our regularization program, started in \cite{ChG1}.  First, we improved our  previous pointwise estimates for the gauge invariant quantities $|B(t)|$ and $|A'(t)|$ as $t\downarrow 0$ whenever $A(t)$ is a strong solution to \eqref{ymh}. Our method  required that the boundary of $M$ be smooth and convex in the sense that its second fundamental form is non-negative, so that the  heat operator of the Neumann Laplacian on functions is bounded.  Our estimates depended on the initial energy of the flow, $\|B_0\|_{L^2}$.

 \begin{theorem}  \label{thmNeu} Let $M$ be a compact 3-manifold with smooth convex boundary. Suppose that  $A(t)$ is a locally bounded strong solution on  $[0,T)$
that satisfies either the Neumann \eqref{N4}, Dirichlet  \eqref{D4} or Marini  \eqref{M2} boundary conditions.  Then there exists $\tau>0$, depending only on $\|B_0\|_2$,
such that
\begin{align*}
\|B(t)\|_\infty &\le 2c_N \|B_0\|_2   \; t^{-3/4},\ \ \text{for} \ \      0 <t  \le 2\tau \ \ \text{and}  \\
\|B(t)\|_\infty &\le 2c_N \|B_0\|_2 \; \tau^{-3/4}, \ \  \text{for} \ \   \tau \le t <\infty.
\end{align*}
Moreover, if $\|A'(0)\|_2 <\infty$ then there exists $\gamma >0$ such that
\[
\|A'(t)\|_\infty  \le \gamma \|A'(0)\|_2 \; t^{-3/4}, \ \  \text{for} \ \   0 < t \le  2\tau .
\]
\end{theorem}

For the proof of this theorem we used the fact that whenever $A(t)$ is a smooth strong solution  to the Yang-Mills heat equation \eqref{ymh} that satisfies Neumann, Dirichlet or Marini boundary conditions, as in the Theorem above, then both its curvature $B(t)$ and $A'(t)$ satisfy a parabolic equation with reasonable potential terms (and with respective classical boundary conditions). Moreover, whenever the boundary is convex, both functions $|B(t)|^2$ and $|A'(t)|^2$  satisfy classical sub-Neumann boundary conditions for any one of the boundary conditions on $A(t)$.

The key element of the proof is the use of a  Neumann domination technique. Namely, over a manifold with convex boundary the heat kernel on forms is dominated by the Neumann heat kernel on functions. For this domination technique, the boundary conditions for the connection and the convexity of the boundary are key. Finally, we can use the ultracontractivity property of the Neumann Laplacian on functions over these manifolds to control $\|B(t)\|_\infty$, $\|A'(t)\|_\infty$ even near $t=0$. Note that the constant $c_N$ that appears in the theorem is
\[
c_N = \sup_{0<t \le 1} t^{3/4} \|e^{t\Delta_N}\|_{2\rightarrow \infty}
\]
where $\|e^{t\Delta_N}\|_{2\rightarrow \infty}$ is the ultracontractivity norm of the Neumann Laplacian on functions.  The constant $\gamma$ depends only on $c_N$ and $\|B_0\|_2.$ After proving that the estimates of Theorem \ref{thmNeu} hold for a smooth solution, we then used the regularization Lemma \ref{lemReg} to prove them for a locally bounded strong solution to \eqref{ymh} with bounded initial energy.

In addition to the small time estimates, we were also interested in the long-time convergence properties of our solutions.  Motivated from the general realization that a gauge invariant regularization method for Wilson loop variables might be necessary for the construction of quantized Yang-Mills fields \cites{Ba,Sei}, we considered the Wilson loop functions in our setting.  To define the Wilson loop function, we first recall that each connection has a parallel transport operator along curves, which in turn determines the connection.  Instead of proving the convergence of the connection itself, we were able to show the convergence of these parallel transport operators. In particular, in \cite{ChG2} we proved that the Wilson loop functions, gauge invariantly regularized, will converge as time goes to infinity for any initial gauge potential $A_0\in H_1$.

\begin{definition}  For a smooth End $\V$ valued
connection form $A$ on the interior of $M$ and a piecewise
$C^1$ path $\gamma: [0,1] \to M$, the parallel transport operator along $\gamma$
is defined by the solution to the ordinary differential equation
\begin{equation*}
g(t)^{-1} dg(t)/dt = A\<  d \gamma(t)/dt \>,\ \ \    g(0) = I_{\V}.
\end{equation*}
We set $//_\gamma^A = g(1)$ and note that this map satisfies the classical properties of a parallel transport (see Notation 3.4 in \cite{ChG2}).

The Wilson loop function is defined as $W_\gamma (A) \equiv trace\ //_\gamma^A$ where
the trace is computed in some finite dimensional unitary representation of $K$.
\end{definition}

We will denote the set of closed loops at a fixed point $x_0$ by
\[
\Gamma_0 = \{ \gamma \;\big|  \ \gamma \text{ is a piecewise} \  C^1 \ \ \text{function}
      \gamma: [0,1] \to M,  \ \text{satisfying} \ \gamma(0) = \gamma(1) = x_0 \}
\]

\begin{theorem} \label{LTB}
Suppose that M is a compact convex subset of $\mathbb{R}^3$ with smooth boundary, and let
$A(\cdot)$ be a locally bounded strong solution of the Yang-Mills heat equation \eqref{ymh} over $[0,\infty)$, satisfying
Dirichlet or Neumann boundary conditions. Choose a point $x_o$ in the interior of $M$. Suppose that $\{t_i\}$ is a sequence of times going to $\infty$. Then, there exists a subsequence $t_j$ and gauge functions $k_j\in  W_1(M; K)$ such that
\begin{enumerate}[$1)$]
\item $k_j^{-1} dk_j \in W_1(M; \mathfrak{k})$  for all $j$

\item  $\alpha_j= A(t_j)^{k_j}$  is in $C^\infty(M;\Lambda^1\otimes\mathfrak{k})$,   and

\item for each $\gamma \in \Gamma_0$,  the operators $//_\gamma^{\alpha_j}$
converge, as operators from $\V$ to $\V$, to a map $P(\gamma)$   as $j\rightarrow \infty$. The map $P$   can be extended to a parallel transport system on the set of loops.
\end{enumerate}
\end{theorem}
For more detailed properties of $P$ and parallel transport systems we refer the interested reader to \cite{ChG2}*{Section III}.

In general  $W_\gamma(A)$ is highly singular as a function of the connection $A$ when $A$ varies over the very large space of typical gauge fields required in quantized theory. As we illustrate in \cite{ChG2} the function $A \mapsto  trace\ //_\gamma^A$ is fully gauge invariant, in the sense that  $trace\ //_\gamma^{A^k} = trace //_\gamma^A$ whenever $\gamma$ is a closed curve in $M^{int}$, $A$ is a smooth connection form and $k$ is a smooth gauge transformation.  At the same time, the Yang-Mills heat equation is itself fully gauge invariant: if we transform the initial data $A_0$ by a gauge transformation  and then propagate, we arrive at the same gauge field as if we first propagate $A_0$ and then gauge transform. Moreover, the flow regularizes the initial data well enough so that the Wilson loop function $W_\gamma(A(t))$ is well-defined for any fixed time $t>0$, even when $W_\gamma(A_0)$ fails to be so, since  $W_\gamma(A(t))$ is gauge invariant. As a result, the Yang-Mills heat equation offers a gauge invariant regularization procedure for the Wilson loop function for some class of irregular connection forms.

Theorem \ref{LTB} implies  that for any initial gauge potential $A_0\in H_1$ there exists a sequence of times going to infinity for which the functions  $\;trace //_\gamma^{A(t_j)}$, in other words $W_\gamma(A(t_j))$,  converge for all piecewise  $C^1$ loops $\gamma$ starting at $x_0$. The proof relies on the fact that the norm of the Wilson loop functions is controlled by the $L^\infty$ norm of the curvature of the connection, which is bounded in this case for $t\geq 1$. The underlying space was  a compact convex subset of $\mathbb{R}^3$ with smooth boundary, since the case of interest for quantum field theory is that in which $M$ is the closure of a bounded open set $O$ in $\mathbb{R}^3$ with smooth boundary.

\section{The Yang-Mills heat equation under finite action} \label{FA}

More recently, Gross has considered the existence and uniqueness of solutions to the Yang-Mills heat equation for less regular  initial data $A_0\in H_{a}(M)$ for $1/2\leq a < 1$ \cite{Gr2}. He considered the case when $M$ is either all of $\mathbb{R}^3$  or the closure of a bounded domain in  $\mathbb{R}^3$  with smooth convex boundary.
 As we have already mentioned, the critical case in dimension 3 is when $a=1/2$, which is the most general case in which one anticipates existence and uniqueness of solutions. In fact, the techniques used in the proof of the existence and uniqueness theorems for  $a>1/2$ break down as $a\downarrow 1/2$ and further illustrate the way in which $a=1/2$ is critical.

Recall that as defined in \eqref{Lap5} the $H_{a}$ norms are not in themselves gauge invariant for $1/2\leq a <1$. One of the central ideas of Gross, was that the functional that does capture in a gauge invariant way the $H_a$ norm of $A_0$ is the following.
\begin{definition}[Finite $a$-action] An almost strong solution of the second type $A(\cdot)$ to the Yang-
Mills heat equation has finite $a$-action if
\begin{equation}\label{fa1a}
\rho_{a}(t)= \int_0^t s^{-a} \|B(s) \|_2^2\, ds < \infty \ \ \ \text{for some} \ \ t>0,
\end{equation}
where $B(s)$ is the curvature of $A(s)$. This definition is of interest for $1/2\leq a<1$ .
\end{definition}
The finite action property for a solution, does control many of the estimates needed in this more general setting. Gross' use of this term was motivated by the observation that when this functional is finite, then the initial condition $A_0$ has an extension to a time interval in Minkowski space with a finite magnetic contribution to the Lagrangian. In the setting of the Yang-Mills heat equation, it allowed Gross to prove the existence and uniqueness of strong solutions of the second type $A(t)$ which belong to $W_1(M)$ for $t>0$, and whose curvature also belongs to  $W_1(M)$ for $t>0$ (see Definition \ref{SS2}). The lack of stronger regularity at $t=0$ is the main difference between strong solutions in the finite action case and strong solutions in the sense of Definition \ref{SS1}. The latter, as anticipated, are no longer possible given the less regular initial value $A_0$. We state below the two main theorems in \cite{Gr2}.

\begin{theorem}[Gross \cite{Gr2}] \label{thmFAa} Let $1/2<a<1$ and assume that $M$ is either all of  $\mathbb{R}^3$  or is the closure of a bounded domain in  $\mathbb{R}^3$  with smooth convex boundary. Suppose that  $ A_0 \in H_{a}(M)$. Then

\begin{enumerate}[$1)$]

\item  there exists an almost strong solution of the second type $A(t)$ to \eqref{ymh} over $[0,\infty)$ with $A(0)=A_0$
which satisfies the following properties.

\item   There exists a gauge function $g_0\in \G_{1+a}$ such that $A(t)^{g_0}$ is a strong  solution of the second type.

\item   $A(\cdot)$ and $A(\cdot)^{g_0}$ are continuous functions on $[0,\infty)$  into $H_a$.

\item   Both  $A(\cdot)$ and $A(\cdot)^{g_0}$ have finite $a$-action.

\item   If $M\neq \mathbb{R}^3$ then the curvature of  both $A(\cdot)$ and $A(\cdot)^{g_0}$ satisfies the boundary condition \eqref{N2}, or respectively \eqref{D2}, for all $t>0$, depending on the boundary condition of $A_0$. Moreover $A(\cdot)^{g_0}$ satisfies the Neumann boundary condition \eqref{N1}, or respectively the Dirichlet  boundary condition \eqref{D1}, for all $t>0$.

\item   Strong solutions of the second type are unique among solutions with finite $a$-action  under the boundary condition \eqref{N2}, respectively \eqref{D1}, for all $t>0$ when $M\neq \mathbb{R}^3$.

\end{enumerate}
\end{theorem}

In other words, any connection form $A_0\in H_{a}$ is, after gauge transformation, the initial value of a strong solution of the second type. Uniqueness holds when properly formulated, and note that since $A(t)$ need not be in $W_1$ in this case, the boundary conditions \eqref{N1} and \eqref{D1} are only meaningful for the gauge transformed solution $A(t)^{g_0}$. A similar result holds true for the case $a=1/2$  however, to gain continuity at $t=0$ into  $H_{1/2}$ and to prove that the solution has finite $(1/2)$-action one must assume that the $H_{1/2}$ norm of $A_0$ is sufficiently small.

\begin{theorem}[Gross \cite{Gr2}] \label{thmFA12} Assume $M$ is either all of  $\mathbb{R}^3$  or is the closure of a bounded domain in  $\mathbb{R}^3$  with smooth convex boundary. Suppose that  $ A_0 \in H_{1/2}(M)$. Then,
\begin{enumerate}[$1)$]
\item   there exists an almost strong solution of the second type $A(t)$ to \eqref{ymh} over $[0,\infty)$  with $A(0)=A_0$. The curvature of $A(t)$ satisfies the boundary condition \eqref{N2}, or respectively \eqref{D2}, for all $t>0$ when $M\neq \mathbb{R}^3$, depending on the boundary condition of $A_0$.

\item There exists a gauge function $g_0$ such that $A(t)^{g_0}$ is a strong  solution of the second type, and  $A(t)^{g_0}$  satisfies  the boundary conditions  \eqref{N1} and \eqref{N2}, or respectively \eqref{D1} and \eqref{D2}, for all $t>0$  when $M\neq \mathbb{R}^3$.

\item If $\; \|A_0\|_{H_{1/2}}$ is sufficiently small, then  $A(\cdot)$ and $A(\cdot)^{g_0}$ have finite $(1/2)$-action. In this case one may choose $g_0 \in \mathcal{G}_{3/2}$.

\item If $\; \|A_0\|_{H_{1/2}}$ is sufficiently small, then  $A(\cdot)$ is a continuous function from $[0,\infty)$ into $H_{1/2}$. If in addition $g_0$ is chosen to lie in $\mathcal{G}_{3/2}$ then $A(\cdot)^{g_0}:[0,\infty) \to H_{1/2}$ is also continuous.

\item Strong solutions are unique among solutions with finite $(1/2)$-action  under the boundary condition \eqref{N2}, respectively \eqref{D1}, for all $t>0$ when $M\neq \mathbb{R}^3$.
\end{enumerate}
\end{theorem}

We note that the gauge transformation $g_0$ in Theorem \ref{thmFAa} that converts an almost strong
solution $A$ to a strong, $A^{g_0}$ one is not unique. In particular, if $g_1$ is an element of ${\mathcal G}_2$, then $A^{g_0 g_1}$ is also a strong solution. It would be interesting to know whether this is the full-extent of non-uniqueness.

As Gross illustrates in \cite{Gr2}*{Theorem 7.1} the solution $A(t)^{g_0}$ produced by the two theorems above is actually in $C^\infty\left( (0,T]\times M; \Lambda^1\otimes \mathfrak{k} \right)$ for some $T<\infty$.  The importance of this property will be illustrated in Theorem \ref{thmEst} and Corollary \ref{corlEst}, where it allowed us to obtain gauge covariant derivatives of all orders and prove improved  $L^p$ and $W_1$ estimates for them for small time \cite{ChG3}.

The proofs of Theorems \ref{thmFAa} and \ref{thmFA12} use similar techniques to the finite energy case, with a lot of technical subtlety and an augmented parabolic equation. The gauge transformation $g_0$ that converts an almost strong solution to  a strong one is unavoidable since, as we have mentioned in the introduction, the solution $A(t)$ can be as irregular as $A(0)$ even if its curvature is smooth. As a result, we cannot expect that any $A_0\in H_{a}$ will be the initial value to a strong solution for $a<1$. This is also reflected in the different way that the ZDS procedure is used in the proof of  Theorem \ref{thmFA12}. If the initial data is in $H_1$, then the gauge transformation flow $g(t)$ produced by the augmented parabolic equation is only used to produce the strong solution $A$ from $C$. But for $A_0\in H_a$ the ZDS procedure produces a gauge transformation $g_0$ such that $A_0^{g_0}$ is the initial value to a strong solution of the second type.

As in the finite energy case, the difficulty in the ZDS procedure arises from the singular behavior of $d^*C(t)$ as $t\downarrow 0$, since $d^*C(0)$ need only belong to $H_{-a}$ in this case.  A significant part of \cite{Gr2} was dedicated to proving that $t\mapsto g(t)$ is a continuous function into $\G_{1+a}$ for $A_0\in H_a$. Gaffney-Friedrichs inequalities in combination with Neumann domination results were the ones that enabled the use of Sobolev inequalities that led to $L^p$ estimates for all $p\leq \infty$. The finite $a$-action condition was key in obtaining many of the $H_a$ estimates. Uniqueness for $a>1/2$ also relied on a Gronwall type argument, but for $a=1/2$ it was necessary to follow a more specialized proof.

Finally, we remark that the notion of solution to the Yang-Mills heat equation in the finite action case with $A_0\in H_{1/2}$ allowed for the first spatial derivatives of  $A(t)$ to exist in some generalized sense. On the other hand, the weak curvature of $A(t)$ is actually in $H_1$ for all $t>0$, and in consequence certain second order derivatives of $A(t)$ can be defined in the classical sense. This is unusual for typical weak solutions in heat equations, but reflects the many symmetries that are satisfied by higher order  derivatives of solutions in the Yang-Mills setting.

In \cite{ChG3} together with Gross we more carefully considered the small-time behavior of solutions to  \eqref{ymh} for initial data $A_0\in H_{1/2}$. We were interested in the smoothness properties of the solution for  $t >0$, and given the fact that in general the higher order covariant derivatives of $A$ itself need not belong to $W_1(M)$, we concentrated only on gauge covariant derivatives. Our main result was the following.

\begin{theorem} \label{thmEst} Assume that $A_0\in H_{1/2}(M; \Lambda^1 \otimes   \frak k )$.
Suppose that $A(\cdot)$ is a strong solution of the second type to \eqref{ymh}
over $[0,\infty)$
with initial value $A_0$ and having finite action. If $\|A_0\|_{H_{1/2}}$ is sufficiently small
then there exists $T>0$ and
standard dominating functions $C_{nj}$  for $\ j=1,\ldots 4$
and  $n=1,2,\dots$,   such that, for $0<t < T$,  the following estimates hold.
\begin{align*}
 t^{2n-\frac 12}\|A^{(n)}(t)\|_2^2 \ +
     &\int_0^t  s^{2n- \frac 12} \|B^{(n)}(s) \|_2^2\, ds  \le C_{n1} (t) \ \ \ \ \         \\
t^{(2n- \frac 12)}\|B^{(n-1)}(t)\|_6^2 \ +
    & \int_0^t  s^{2n- \frac 12} \|A^{(n)}(s) \|_6^2 \,ds  \le C_{n2}(t)           \\
t^{2n+\frac 12} \| B^{(n)} ( t)\|_2^2 \ +
     & \int_0^t  s^{2n+\frac 12} \| A^{(n+1)} (s) \|_2^2 \,ds  \le  C_{n3} (t)         \\
t^{2n+\frac 12} \| A^{(n)} (t)\|_6^2  \ +
     & \int_0^t  s^{2n+\frac 12} \| B^{(n)}(s)\|_6^2 \,ds \le C_{n4}(t).
\end{align*}
Moreover the third estimate also holds for $n =0$.
\end{theorem}

In the above theorem $A^{(n)}$ and $B^{(n)}$ denote the $n$th order time-derivatives of $A$ and $B$ respectively. A  standard dominating function is a function $C:[0, \infty) \rightarrow [0,\infty)$ of the form $C(t) =\hat C(t, \rho_{1/2}(t))$ , where $\rho_{1/2}(t)$ is the finite $1/2$-action functional at time $t$, such that  $\hat C:[0,\infty)^2 \rightarrow [0, \infty)$ is continuous and non-decreasing   in each variable, $\hat C(0,0) =0$ and $\hat C$ is independent of the solution $A(\cdot)$.

The  estimates of Theorem \ref{thmEst} provided information about the order of the singularity as time $t\downarrow 0$ which are consistent with what is expected in parabolic equations for the respective order of the derivative and for initial data in the  Sobolev space  $H_{1/2}$. At the same time, estimates for $A^{(n)}$ and $B^{(n)}$ are essentially estimates for higher order covariant exterior derivatives and coderivatives of $A$ and $B$. For example, we know that $A'(t)=-d^*_{A(t)} B(t)$ and $B'(t)= d_{A(t)} A'(t)$  therefore our small-time estimates are in fact $L^2$ and $L^6$ estimates for first order spacial derivatives of $A$ and $B$. In \cite{ChG3} the identities we proved for $A^{(n)}$ and $B^{(n)}$ would also imply $L^p$ estimates for higher order spacial derivatives of $A$ and $B$.

A central idea behind the proof of the theorem was to use  Gross' result in \cite{Gr2} which states that a strong solution of the second type $A$ to \eqref{ymh} with $A_0\in H_{1/2}$ and $\|A_0\|_{H_{1/2}}$ small enough is gauge equivalent to a smooth solution $\hat{A}=A^{g_0}$.  The smooth solution  $\hat{A}$ exists for small time, satisfies the same (respective) boundary conditions and also has finite $(1/2)$-action.  At the same time, all $n$th order time derivatives $\hat{A}^{(n)}(t)$ for $n\geq 1$,  and $\hat{B}^{(n)}(t)$ for $n\geq 0$  are well defined for the smooth solution, and in addition,  their norms are gauge invariant quantities.    As a result, all quantities on the left side of the inequalities of Theorem  \ref{thmEst} are gauge invariant and since the estimates hold for  $\hat{A}$, they will automatically hold for the original solution $A$.

The remaining proof consisted in showing that the gauge covariant  exterior derivatives
and coderivatives of $A^{(n)}(t)$ and $B^{(n)}(t)$ (for the smooth solution) can be expressed in terms of lower order time derivatives. These differential identities then led to integral identities for the $L^p$ norms of these quantities which in turn were used to establish bounds on the initial behavior by induction on $n$. This was similar to the process described in Subsection \ref{Apr}, and took advantage of the many symmetries that characterize higher order derivatives of solutions to \eqref{ymh}. The proof made an extensive use of the Gaffney-Friedrichs inequality of Theorem \ref{thmGF} and the Sobolev embedding \eqref{eqSob}.  Particular care had to be taken in the case that the boundary of the manifold was nonempty, so that the correct boundary conditions would hold for all quantities.

The proof of the theorem also led to short-time estimates for the $H_1$ norm of the higher order time-derivatives of $A$ and $B$.
\begin{corollary}\label{corlEst} Under the hypotheses of Theorem \ref{thmEst} there exists $T>0$
 and standard dominating functions $C_{nj}$   for $j=5,6$  and $n=1,2,...$  such that, for  $0 < t <T$,  the following estimates hold.
\begin{align*}
 \qquad t^{(2n- \frac 12)}\|B^{(n-1)}(t)\|_{H_1^A}^2 + & \int_0^t
             s^{2n- \frac 12} \|A^{(n)}(s) \|_{H_1^A}^2 \,ds              \le C_{n5}(t)
             \\
 \qquad\ \ \  \  \ t^{2n+\frac 12} \| A^{(n)} (t)\|_{H_1^A}^2  + & \int_0^t
            s^{2n+\frac 12} \| B^{(n)}(s)\|_{H_1^A}^2 \,ds                   \le C_{n6}(t).
\end{align*}
\end{corollary}

In this setting we make the following interesting observation. Let $\mathcal{Y}$ denote the set of almost strong solutions to the Yang-Mills heat equation over $M$ with initial value   $A_0\in H_{1/2}$ and having finite action.  Theorem \ref{thmFA12} tells us that the group $\G_{3/2}$ acts on $\mathcal{Y}$ through its action on $A(0)$ for each $A \in \mathcal{Y}$.   Since all functionals that appear on the both sides of the estimates in Theorem \ref{thmEst} and Corollary \ref{corlEst} are gauge invariant, then they all descend to  functions  of the initial values on  the quotient space ${\mathcal{C}} \equiv \mathcal{Y}/\G_{3/2}$. In other words, our estimates are in fact estimates on $\mathcal{Y}/\G_{3/2}$.

\section{The configuration space for the Yang-Mills heat equation} \label{S4}

Relating these results to the main questions from Quantum Field theory, it is natural to ask what would be a well-defined configuration space for the Yang-Mills heat equation. A configuration space for classical Yang-Mills fields is generally defined as a quotient space $\mathcal{C} = \mathcal{Y}/\mathcal{G}$ where $\mathcal{Y}$ is an appropriately chosen space of connections and $\mathcal{G}$ is an appropriate group of gauge transformations. The structure of the configuration space for classical Yang-Mills Fields. remains a central but still elusive problem in Mathematical Physics. At the same time, in order to carry out quantization for the classical Yang-Mills field, in other words assign a metric or measure structure to the configuration space, it is important to choose $\mathcal{Y}$ and  $\mathcal{G}$ such that the quotient space is a complete metric space in a natural metric and in particular a Hilbert manifold. For these structures it is appropriate to start this process over a compact subset of $\mathbb{R}^3$ and then extend to the whole space, which also motivated our study of compact 3-manifolds with boundary.

Gross anticipates that in this setting if $\mathcal{Y}_a$ is the space of strong solutions to the Yang-Mills heat equation over $\mathbb{R}^3$ with finite $a$-action and initial value   $A_0\in H_{a}(\mathbb{R}^3)$, then $\mathcal{Y}_a/\G_{1+a}$ is complete metric space, and in fact it is a Hilbert manifold for $1/2< a<1$ \cite{Gr4}. A similar result should also hold for $a=1/2$. See also \cite{Gr1} for a notion of configuration space for Yang-Mills fields in the context of the Maxwell-Poisson equation.

To this end, he has recently worked on defining an appropriate `tangent space' to each solution to the Yang-Mills heat equation. The notion of a tangent space to a solution is similar to the one from differential geometry. Here one considers paths $[r_1,r_2] \ni r \mapsto A_r(\cdot)$ where $ A_r(\cdot)$ is a solution to \eqref{ymh} with $A_r(0)\in H_{a}$. If $v_r(s) = \partial_r A_r(s)$ in the $L^2$ sense for each $r$, then $v_r$ gives the analogue of a tangent vector to $A_r$. Moreover, these tangent vectors must be solutions to the variational equation
\begin{equation} \label{veq}
-v'(t)= d^*_{A(t)} d_{A(t)} v(t) + [v(t) \lrc B(t)].
\end{equation}

In a recent preprint  Gross proved the existence  of solutions to the variational equation for initial conditions $v_o \in H_{a}(M)$ when $A(t)$ is a strong solution to \eqref{ymh} with finite action \cite{Gr3}. He considered only the case where $M$ is either all of $\mathbb{R}^3$ or a bounded subset of it with smooth convex boundary. In this context, the definition of strong solution is slightly more general than Definition \ref{SS2}.
\begin{definition} \label{SS3}  A {\bf \it  strong solution of the third type} to the Yang-Mills heat equation over $(0, \infty)$
is a continuous function
\begin{equation*}
A(\cdot): (0,\infty) \rightarrow  L^2(M; \Lambda^1 \otimes   \frak k )
\end{equation*}
which satisfies the conditions $a)-d)$ of Definition \ref{SS2} and in the case the boundary of the manifold $M$ is nonempty $A$ is assumed to satisfy the boundary conditions $A(t)_{norm}=0$ in the Neumann case and $A(t)_{tan}=0$ in the Dirichlet case.
\end{definition}
In particular, the strong solutions $A(t)$ for which the variational equation is defined, need not have initial condition in $H_a$, nor any type of continuity at $t=0$.

\begin{definition} \label{SSV}   A {\bf \it  strong solution} to the variational equation \eqref{veq} over $[0, \infty)$
is a continuous function
\begin{equation*}
v : [0,\infty) \rightarrow  L^2(M; \Lambda^1 \otimes   \frak k )
\end{equation*}
such that
\begin{align}
a)& \ v(t) \in H_1^{\textup{A}} \  \text{for all}\ \  t\in (0,\infty),\    \text{and}   \ \  v : (0,\infty) \rightarrow  H_1^{\textup{A}}  \  \text{is continuous} \notag     \\
b)& \ d_{A(t)}v(t) \in H_1^{\textup{A}} \  \text{for each}\ \  t\in (0,\infty),\   \notag     \\
c)& \ \text{the strong $L^2(M)$ derivative $v'(t) \equiv dv(t)/dt $}\
                                      \text{exists on}\ (0,\infty),   \text{and}       \notag \\
d)& \    \text{the variational equation \eqref{veq} holds on}\  (0, \infty).        \notag
\end{align}
A solution $v(\cdot)$ that satisfies all of the above conditions
 except for $a)$ will be called an {\it almost strong solution}. In this
 case the spatial exterior derivative $dv(t)$  must  be interpreted in the weak sense.
\end{definition}

In the above definition, the Sobolev norm $H_1^{\textup{A}}$ is defined as
\[
\| \w \|_{H_1^{\textup{A}}(M)}^2
 = \int_M |\p_j^{\textup{A}} \w(x)|_{ \Lambda^1\otimes\frak k}^2 + | \w(x)|_{ \Lambda^1\otimes\frak k}^2 d\, x\ \
                                        + \| \w \|_2^2,
 \]
since we are over Euclidean space, and $\textup{\rm A}=A(T)$ for some $0<T<\infty$. The $H_a^\textup{\rm A}$ norm is defined similarly to \eqref{Lap5}, with $\Delta$ replaced by $\Delta_\textup{\rm A}$.

\begin{theorem}[Gross \cite{Gr3}] \label{thmVE} Assume $M$ is either all of  $\mathbb{R}^3$  or is the closure of a bounded domain in  $\mathbb{R}^3$  with smooth convex boundary. Assume that $1/2\leq a<1$ and $1/2\leq b<1$. Let  $A(\cdot)$ be a strong solution to the Yang-Mills heat equation over $(0,\infty)$ with finite
$a$-action and such that for each $s\in [0,\infty)$ the function
\[
[0,\infty) \ni  t \mapsto A(t)- A(s) \ \ \text{is continuous into} \  L^3(M; \Lambda^1\otimes\frak k).
\]
Let $v_0 \in H_b^\textup{\rm A}(M;\Lambda^1\otimes\frak k)$. Then
\begin{enumerate}[$1)$]
  \item There exists an almost strong solution $v(\cdot)$ to the variational equation
\eqref{veq} over $[0,\infty)$ with initial value $v_0$.
  \item  For each real number $\tau > 0$ there exists a vertical almost strong solution
$d_{A(t)}\alpha_\tau$ for some $\alpha_\tau \in H_1^{\textup{A}}(M;\frak k)$ such that the function
\[
v_\tau (t)  \equiv v(t)  - d_{A(t)}\alpha_\tau, \ t \geq 0
\]
is a strong solution to the variational equation with initial value  $v_0 - d_{A(0)}\alpha_\tau$.
Moreover
\[
\sup_{0\leq t\leq 1} \| v(t) - v_\tau (t)\|_2 \to 0 \ \ \text{as} \ \ \tau \downarrow 0.
\]
\item If $\|A(t)\|_{L^3(M)} < \infty $ for some $t > 0$  then
\[
v : [0,\infty) \to  H_b^\textup{\rm A}
\]
is continuous.
\item Strong solutions are unique when they exist.
\end{enumerate}
\end{theorem}
Theorem \ref{thmVE} implies that the solution satisfies   $d_{A(t)} v(t) \in H_1,$ but fails to be in $H_1$ up to a vertical solution. In other words, as Gross mentions in \cite{Gr3}, the above result is the infinitesimal analogue of the existence theorem for the Yang-Mills heat equation, where now the infinitesimal analogue of a gauge transformation is played by the vertical vectors. To achieve $v\in  H_b^\textup{\rm A}$, one must make the additional assumption that $A$ is in $L^3$. This is known for initial data $A_0\in H_{1/2}$, but need not hold in general

Recently, Gross in \cite{Gr4}, and in a separate work with the author, have been considering the topological properties of the natural configuration space that arises in the context of the Yang-Mills heat equation over compact subsets of 3-dimensional Euclidean space with smooth boundary and $\mathbb{R}^3$ itself. They are interested in providing a space of solutions $\mathcal{Y}$ to  \eqref{ymh} as well as an appropriate space of gauge transformations $\G$ such that $\mathcal{Y}/\mathcal{G}$ is an infinite dimensional complete manifold, with a metric structure. The spaces $\mathcal{Y}$ and $\mathcal{G}$  should correspond to a general class of initial conditions, and in particular one that would be relevant for quantum field theory applications. They anticipate that the space of solutions with initial value in $A_0 \in H_{1/2}(M)$, or alternatively $A_0 \in H_{a}(M)$,  and an appropriate gauge group will provide such a space. The previous work of Gross regarding the solutions to the variational equation will be critical, as the metric will correspond to a norm on the relevant tangent vectors, which will eventually allow us to find a homeomorphism between small neighborhoods of the space and the relevant tangent space.

\end{document}